# 5G New Radio Evolution Meets Satellite Communications: Opportunities, Challenges, and Solutions

Xingqin Lin, Björn Hofström, Eric Wang, Gino Masini, Helka-Liina Maattanen, Henrik Rydén, Jonas Sedin, Magnus Stattin, Olof Liberg, Sebastian Euler, Siva Muruganathan, Stefan Eriksson G., and Talha Khan

Ericsson

Contact: xingqin.lin@ericsson.com

*Abstract*— **The 3rd generation partnership project (3GPP) completed the first global 5th generation (5G) new radio (NR) standard in its Release 15, paving the way for making 5G a commercial reality. So, what is next in NR evolution to further expand the 5G ecosystem? Enabling 5G NR to support satellite communications is one direction under exploration in 3GPP. There has been a resurgence of interest in providing connectivity from space, stimulated by technology advancement and demand for ubiquitous connectivity services. The on-going evolution of 5G standards provides a unique opportunity to revisit satellite communications. In this article, we provide an overview of use cases and a primer on satellite communications. We identify key technical challenges faced by 5G NR evolution for satellite communications and give some preliminary ideas for how to overcome them.**

## I. INTRODUCTION

The 5th generation (5G) wireless access technology, known as new radio (NR), features spectrum flexibility, ultra-lean design, forward compatibility, low latency support, and advanced antenna technologies [1]. Built on the first release of NR (Release 15), the evolution of NR will bring additional capabilities to provide better performance and address new application areas. Meanwhile, we are witnessing a resurgent interest in providing connectivity from space. In the past few years, there has been a surge of proposals about using large constellations of low earth orbit (LEO) satellites, such as *OneWeb* [2] and *SpaceX* [3], to provide broadband access, in addition to the existing satellite communications systems such as Iridium. It is anticipated that the integration of satellite communications into 5G will facilitate anything, anytime, anywhere connectivity in the 5G era and beyond.

The ambition of providing connectivity from space is not new. A series of satellite communications projects (e.g., Iridium and Globalstar) were planned in the 1990's but with limited success, partly due to the fast growth of terrestrial networks that were more economically appealing. A resurgence of interest in providing connectivity from space started around 2014, stimulated by technology advancement and demand for ubiquitous connectivity services. The advancement of microelectronics following Moore's law has paved the way for using advanced technologies in satellite communications such as multi-spot beam technologies, onboard digital processing, and advanced modulation and coding schemes [4]. Meanwhile, the development cycle and the costs of satellite manufacturing and launching processes have been dramatically reduced [5].

A major driver of the success of terrestrial mobile networks over the past few decades has been the international standardization effort yielding the benefits of significant economies of scale. The 3rd generation partnership project (3GPP) has been the dominating standardization development body of several generations of mobile technology. The international standardization effort helps ensure compatibility among vendors and reduce network operation and device costs. In contrast, the interoperability between different satellite solution vendors has been difficult and the availability of devices is limited, leading to an overall fragmented satellite communications market up to date [6].

The satellite industry has realized the need to embrace standardization, and furthermore to join forces with the mobile industry in 3GPP. The on-going evolution of 5G standards provides a unique opportunity to revisit satellite communications. The satellite industry has been active in the 3GPP 5G standards process. Thus far, 3GPP has completed a few study items on 5G evolution for satellite communications [7][8] and is currently conducting further studies. In addition to the efforts in industry, there have been contributions to the field of satellite communications from academia [9] [10]. We refer interested readers to the recent *IEEE JSAC* special issue for more theoretical works on satellite communications [11].

The objective of this article is to investigate the opportunities and challenges associated with adapting 5G NR for satellite communications. A similar effort has been made for adapting the 4th generation (4G) wireless access technology – long-term evolution (LTE) – for satellite communications [12]. There are a few other works such as [13] and [14] that touched upon satellite communications for 5G, while our article further address NR specific aspects. Another related work [15] examined the impact of satellite channel characteristics on NR physical and medium access control layers, but the analysis was based on NR specifications released in December 2017, which only support non-standalone NR deployment and have been significantly updated by 3GPP in 2018.

## II. USE CASES OF SATELLITE COMMUNICATIONS

Satellite access networks have been playing a complementary role in the communications ecosystem. Despite the wide deployment of terrestrial mobile networks, there are unserved or underserved areas around the globe due to economic rationales. For example, providing coverage in rural or remote areas has been challenging in many countries because the investment cost may not justify the expected revenue. In contrast, a single communication satellite can cover a large geographic area, and thus it might be economically appealing to use satellite communications to augment terrestrial networks to provide connectivity in rural and remote areas. In urban areas, high-throughput satellites communications systems may help offload traffic in terrestrial networks. Another potential alternative is to use satellites for backhauling, fostering the rollout of 5G services with potentially reduced costs in rural and remote areas.

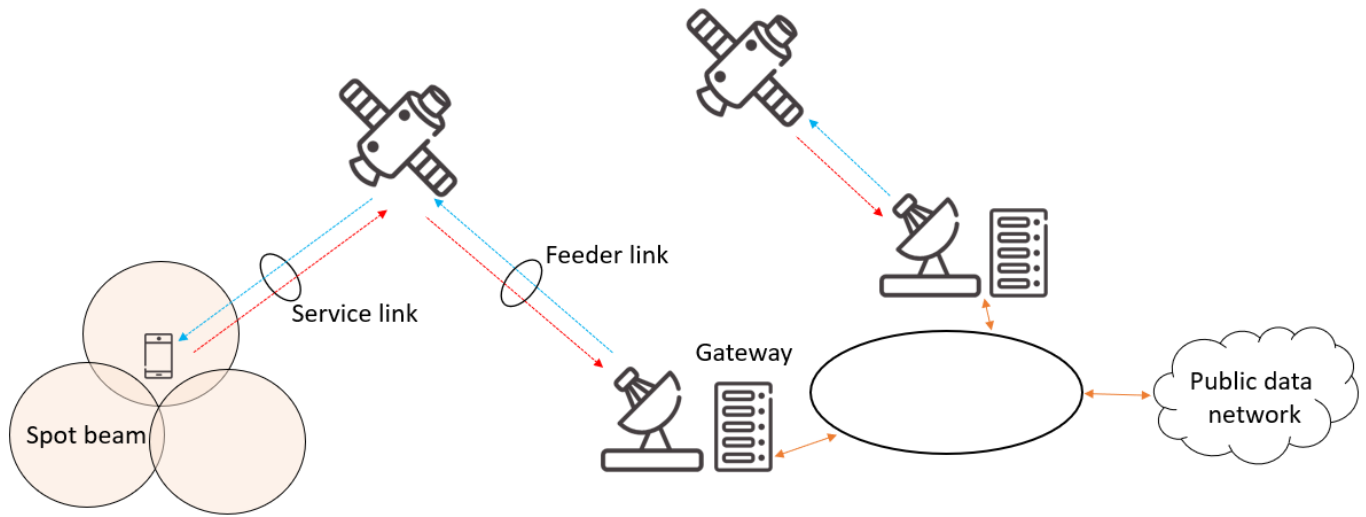

**Figure 1: An illustration of satellite communications system architecture**

The large satellite coverage can also benefit communication scenarios with airborne and maritime platforms (onboard aircrafts or vessels), while being attractive in certain machine-to-machine and telemetry applications. Additionally, satellites are resilient to natural disasters on earth, making satellite communications key for emergency services in case that the terrestrial network infrastructures are degraded.

Service continuity is closely related to ubiquitous connectivity. When a user equipment (UE) enters an unserved or underserved area, the connectivity service may be disrupted. Integrating satellite communications into terrestrial networks to fill the coverage holes can enable smoother service continuity.

Satellite communications are well positioned for broadcasting/multicasting data and media to a broad audience spread over a large geographical area. While television broadcasting has undoubtedly been the main satellite service in this area, there are other use cases. For instance, mobile operators and Internet service providers can also utilize satellite communications to multicast content to the network edge to facilitate content caching for local distribution.

## III. PRELIMINARIES OF SATELLITE COMMUNICATIONS

In this section, we describe the basics of satellite communications.

### A. Satellite Orbits

The height of a satellite orbit, which is the distance of the satellite from the earth's surface, determines the satellite orbital speed around the earth. The satellite's orbit also depends on the orbital eccentricity and inclination in addition to the height. There are three main types of satellites based on the orbital heights.

*Geosynchronous/geostationary earth orbit (GEO):* A geosynchronous satellite has an orbital height of 35,786 km, and its orbit matches the earth's rotation with an orbital period of 24 hours. From a ground observer's perspective, the satellite appears fixed at a single longitude, though it may drift north and south. If the orbital eccentricity and inclination are both zero, the satellite orbits over the equator in a circular orbit. The resulting orbit is geostationary, and the satellite appears to be in a fixed position to an observer on the ground.

*Medium earth orbit (MEO):* The orbital height of MEO satellites ranges from 2,000 km above the earth's surface up to the height of a GEO satellite. The associated orbital period ranges from about 2 to 24 hours.

*Low earth orbit (LEO):* The orbital height of LEO satellites is mainly from 400 km to 2,000 km above the earth surface. LEO satellites move fast and rotate around the earth every 1.5 to 2 hours. Since a LEO satellite quickly changes its position relative to an observer on the ground, the satellite is only visible to the observer for a few minutes.

### B. Satellite Communications System Architecture

Generally, a satellite communications system consists of the following components [7], with an illustration given in Figure 1.

- Satellite: Spaceborne platform including spacecraft bus for satellite operation (power, thermal control, altitude control, etc.) and communication payload (antennas and transponders);
- Terminal: UE used by an end-user to communicate with the network;
- Gateway: Ground station for connecting a satellite to other parts of the network;

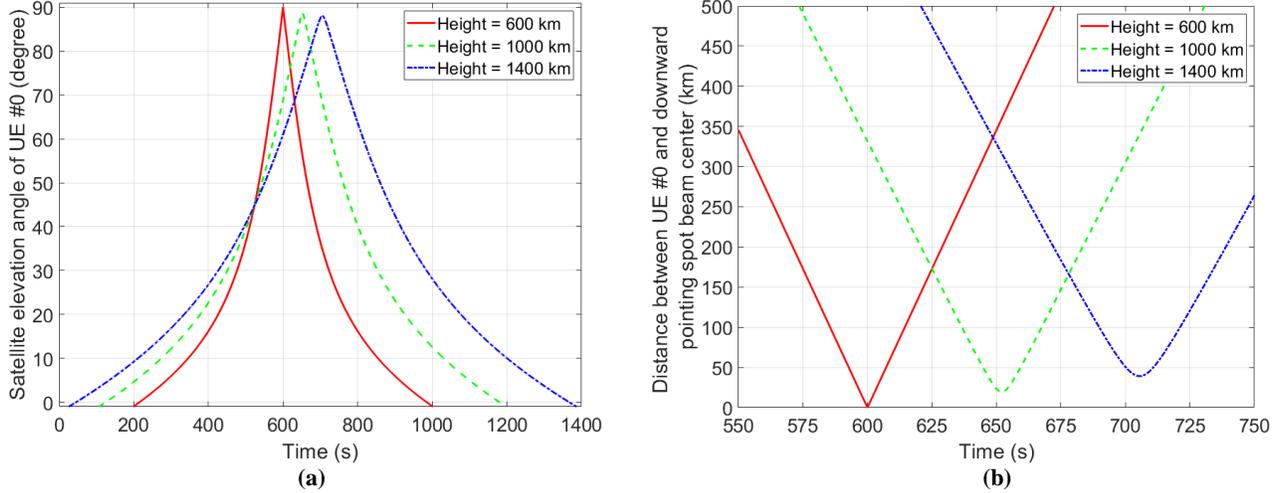

**Figure 2: Varying coverage in satellite communications with polar orbiting satellites at three different orbital heights: subfigure (a) shows satellite elevation angle trajectories of a static reference UE #0 as a function of time; subfigure (b) shows the trajectories of the distance between the reference UE #0 and the center of downward pointing spot beam as a function of time.**

- Feeder link: The communication link between the satellite and the gateway;
- Service link: The communication link between the satellite and the terminal.

Depending on the implemented functionality of the communication payload of the satellite in the system, we can consider two payload options: *bent-pipe* transponder and *regenerative* transponder. With a bent-pipe transponder, the satellite receives uplink signals from the earth, amplifies the received signals, and retransmits the signals to the earth with uplink-downlink frequency conversion. With a regenerative transponder, the satellite performs onboard processing to demodulate and decode the received uplink signals and regenerates the signals for further transmission.

A modern satellite typically uses multi-spot beam technology to generate multiple high-power beams to cover a geographical area. The footprint of a beam, often referred to as spot beam, is usually in an elliptic shape, and is commonly considered as equivalent to a cell in a terrestrial network. For a non-geostationary satellite, the footprint may sweep over the earth's surface with the satellite movement or may be earth fixed with some beam pointing mechanism used by the satellite to compensate for its motion. The radii of spot beams depend on the satellite communications system design, and may range from tens of kilometers to a few thousands of kilometers.

## IV. SALIENT CHARACTERISTICS OF SATELLITE COMMUNICATIONS

In this section, we describe the salient characteristics of satellite communications that heavily impact 5G NR adaptations for satellite communications.

### A. Varying Coverage in Time and Space

The coverage of a GEO satellite is quite static, with infrequent updates of spot beam pointing directions to compensate for the GEO satellite movement in order to have the same spot beam cover the same geographical area. In contrast, the movements of non-GEO satellites, especially LEO satellites, lead to a varying coverage in time and space. A typical LEO satellite is visible to a ground UE for a few minutes only. This implies that even in a LEO satellite communications system with earth fixed beams, where each LEO satellite constantly updates its beam pointing directions to serve a certain geographical area, the serving satellites change every few minutes. In a LEO satellite communications system with moving beams, a typical spot beam with a radius of tens of kilometers can cover a UE for only a few seconds. The varying coverage in time and space clearly has implications for UE mobility management methods when adapting 5G NR for non-GEO satellite communications.

Figure 2 gives an illustration of the varying coverage in LEO satellite communications with polar orbiting satellites at three different heights: 600 km, 1000 km, and 1400 km. Figure 2 (a) shows satellite elevation angle trajectories of a static reference UE #0 as a function of time. Assuming a typical 10° minimum satellite elevation angle for service link connection, the UE can stay connected with the satellite passing at 600 km height above for only about 450 s. Figure 2 (b) shows the trajectories of the distance between the reference UE #0 and the center of downward pointing spot beam as a function of time. If the spot beam radius is 50 km, the spot beam from the satellite at the height of 600 km covers the UE for only about 15 s.

### B. Propagation Delays

Rapid interactions between a UE and its serving base station in a terrestrial mobile communications system are possible since the propagation delay is usually within 1 ms. In contrast, the propagation delay in a satellite link is much longer. The one-way propagation time between a GEO satellite and a ground UE is 119.3 ms, assuming that the radio signal propagates at the speed of light in a vacuum and that the UE is immediately underneath the GEO satellite.

The propagation delay in LEO systems is much shorter than in GEO systems. Figure 3 (a) shows service link propagation delay trajectories of a static reference UE #0 as a function of time at three orbital heights. With 600 km LEO satellite height, the minimum service link propagation delay is 2 ms attained at 90° satellite elevation angle, which increases to 6.5 ms at 10° satellite elevation angle. Accordingly, the minimum round-trip propagation delay for a signal passing between a ground station and a UE in the LEO system is 8 ms. The round-trip propagation delay can increase to about 28 ms assuming 5° elevation angle for the ground station and 10° elevation angle for the UE. In short, the propagation delays are much larger than the typical propagation delays encountered in terrestrial mobile systems.

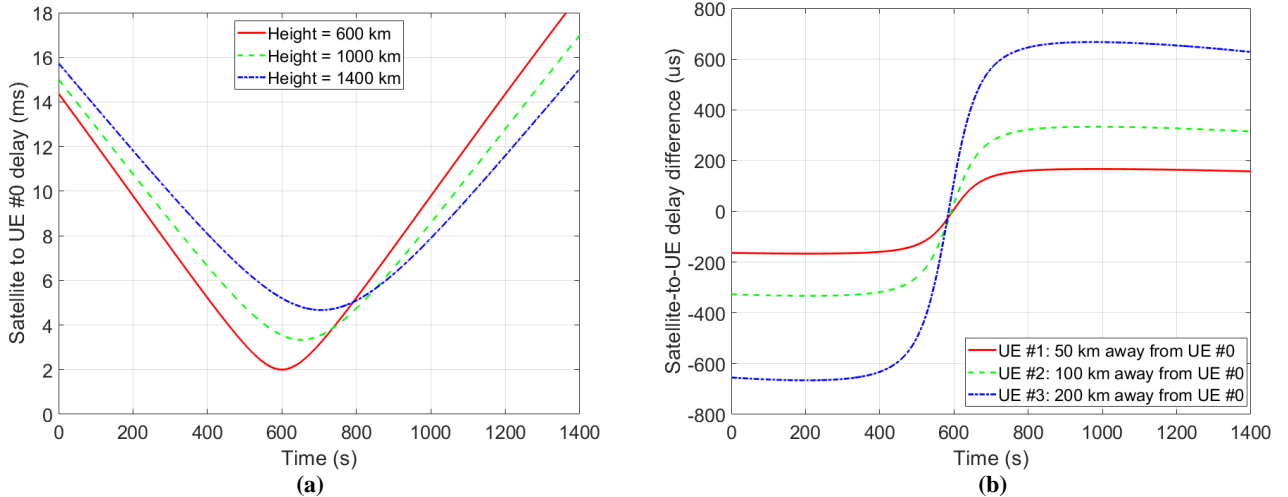

**Figure 3:** Propagation delays in satellite communications with polar orbiting satellites: subfigure (a) shows service link propagation delay trajectories of a static reference UE #0 as a function of time at three orbital heights; subfigure (b) shows the trajectories of service link propagation delay difference of reference UE #0 and UE #i, i=1, 2, 3, as a function of time at 600 km orbital height.

In addition to absolute propagation delays, the differential delay, which refers to the propagation delay difference of two selected points in the same spot beam, is of interest as it impacts the multi-access scheme. Since the feeder link is shared by the devices in the same spot beam, the differential delay mainly depends on the size of the spot beam resulting in different path lengths of the service links. Figure 3 (b) shows the trajectories of service link propagation delay difference of reference UE #0 and UE #i, i=1, 2, 3. Consider a LEO satellite system with earth fixed beams and assume that the location of UE #0 is chosen as the center of a spot beam. Then the curves corresponding to UE #1, UE #2, and UE #3 in Figure 3 (b) indicate the maximum delay difference in the cell if the radius of the spot beam is chosen to be 50 km, 100 km, and 200 km, respectively. Clearly, the larger the spot beam radius, the larger the maximum differential delay in the spot beam, as illustrated in the figure.

### C. Doppler Effects

Doppler effect refers to the change of frequency of a wave due to the movements of the source, observer, and/or objects in the propagation environment. It depends on the relative speed of motion and the carrier frequency. In terrestrial mobile communications systems, Doppler effects are typically caused by the movements of the UE and surrounding objects, while in satellite systems the satellite movement induces additional Doppler effects.

Doppler effect is quite pronounced in LEO systems. At the height of 600 km, a LEO satellite moves at the speed of 7.56 km/s, which can result in a Doppler shift value as large as about 48 kHz at the carrier frequency of 2 GHz, as illustrated in Figure 4 (a). In addition, Figure 4 (a) shows that the Doppler shift value varies rapidly over time, and the rate of such variation is referred to as the Doppler variation rate. To cope with the pronounced Doppler effects, Doppler compensation techniques need to be implemented.

The Doppler effects due to satellite movements in GEO systems in most cases can be negligible. Note that when a satellite is in near GEO orbit with inclination up to 6°, the Doppler shift can reach around 300 Hz at the carrier frequency of 2 GHz [7], but terrestrial mobile technologies such as 5G NR have been designed to handle this order of magnitude of Doppler shift values. For a satellite communications system operating at higher frequency, the Doppler shift increases but this can be handled by 5G NR using scalable numerology.

## V. DESIGN ASPECTS

In this section, we describe the key areas that require adaptation to evolve 5G NR for satellite communications.

### A. Uplink Timing Control

5G NR utilizes orthogonal frequency-division multiple access (OFDMA) as the multi-access scheme in the uplink. The transmissions from different UEs in a cell are time-aligned at the 5G NodeB (gNB) to maintain uplink orthogonality. Time alignment is achieved by using different timing advance values at different UEs to compensate for their different propagation delays. The required timing advance for a UE is roughly equal to the round-trip delay between the UE and gNB.

For the initial timing advance, after a UE has synchronized in the downlink and acquired certain system information, the UE transmits a random-access preamble on physical random-access channel (PRACH). The gNB estimates the uplink timing from the received random-access preamble and responds with a timing advance command. This allows the establishment of initial timing advance for the UE.

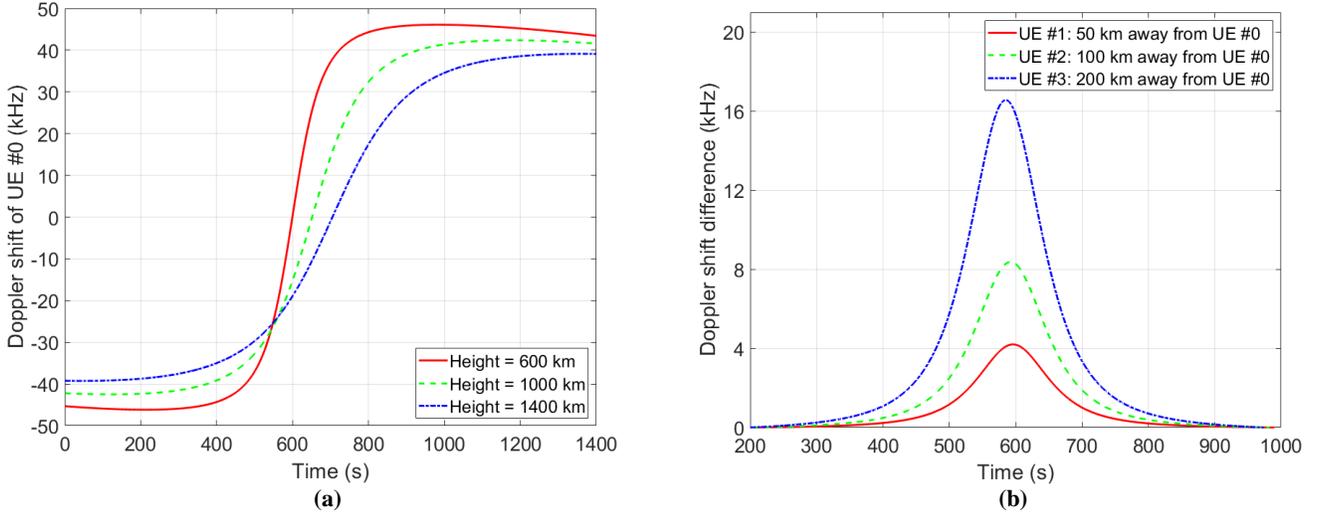

**Figure 4: Doppler effects in satellite communications with polar orbiting satellites and carrier frequency of 2 GHz: subfigure (a) shows service link Doppler shift trajectories of a static reference UE #0 as a function of time at three orbital heights; subfigure (b) shows the trajectories of service link Doppler shift difference of reference UE #0 and static UE #i, i=1, 2, 3, as a function of time at 600 km orbital height.**

The longest cyclic prefix length of NR PRACH formats is about 1.37 ms, and thus may be used for large cells with radii up to about 200 km. This is clearly not enough to handle the much larger propagation delays incurred in satellite communications. One promising approach is to rely on global navigation satellite system (GNSS) based techniques. Each UE equipped with a GNSS chipset determines its position, calculates its propagation delay with respect to the serving satellite using ephemeris data of the satellite constellation, and derives the initial timing advance value. The UE then uses its initial timing advance value to initiate the random-access procedure, which can help to further refine the timing advance to cope with residual timing error.

Some low-cost, reduced complexity UEs may not be equipped with GNSS chipsets. Thus, non-GNSS based techniques are also needed. One possible technique may work as follows. For each spot beam, the gNB may choose a reference point such as the center of the spot beam, and thus may adjust its uplink receiver timing with respect to the reference point. With this approach, the uplink timing control only needs to handle the delay difference between each UE and the reference point instead of the much larger absolute propagation delays. The existing uplink timing control can be directly used for spot beams with radii up to about 200 km. For spot beams with radii larger than 200 km, further adaptation of uplink timing control design may be needed.

### B. Frequency Synchronization

5G NR uses orthogonal frequency division multiplexing (OFDM) for both downlink and uplink transmissions, and additionally supports the use of discrete Fourier transform (DFT) spread OFDM (DFT-S-OFDM) in the uplink. Maintaining the orthogonality of OFDM requires tight frequency synchronization between transmitter and receiver to avoid inter-subcarrier interference. A NR UE can make use of specially designed physical synchronization signals to achieve time and frequency synchronization in the downlink. In the uplink, the UE adds a shift to its acquired downlink reference frequency to obtain the uplink reference frequency based on the known relationship of uplink-downlink carrier frequencies.

The downlink synchronization can be treated as a point-to-point OFDM synchronization problem since each receiver in a cell tunes its downlink reference frequency based on the received synchronization signals. The uplink synchronization is more challenging since it is a multipoint-to-point synchronization problem in OFDMA based 5G NR. The transmissions from different UEs in a cell need to be frequency-aligned at the gNB to maintain uplink orthogonality. Therefore, different frequency adjustment values at different UEs are needed in the uplink to compensate for their different Doppler shifts. GNSS based techniques can be used for uplink frequency adjustment: Each UE equipped with a GNSS chipset determines its position and calculates its frequency adjustment value using its position information, satellite ephemeris data, and carrier frequencies.

To mitigate the effects of large Doppler shifts due to satellite movements in non-GEO satellite communications systems, pre-compensation can be applied to forward link signals: A time-varying frequency offset tracking the Doppler shift is applied to the forward link reference frequency such that the forward link signals for a spot beam received at a reference point in the spot beam appear to have zero Doppler shift. With pre-compensation, the Doppler shift of the forward signals received at a given location in the spot beam becomes equal to the difference between the original Doppler shifts of the given location and the reference point. The Doppler shift differences at different locations in the spot beam however are different and time-varying.

Figure 4 (b) shows the trajectories of service link Doppler shift difference of a static reference UE #0 and UE #i, i=1, 2, 3, as a function of time at 600 km orbital height. Consider a LEO satellite system with earth fixed beams and assume that the location of UE #0 is chosen as the center of a spot beam and frequency pre-compensation is applied to the downlink of the service link with respect to the center of the spot beam. Then the curves corresponding to UE #1, UE #2, and UE #3 in Figure 4 (b) indicate the maximum Doppler shift difference trajectories in the cell if the radius of the spot beam is chosen to be 50 km, 100 km, and 200 km, respectively. Clearly, the larger the spot beam radius, the larger the maximum Doppler shift difference in the spot beam, as illustrated in the figure. As an example, for the spot beam with 100 km radius, the Doppler shift difference of a point at the edge of the spot beam and a reference point in the center of the spot beam can still be as large as 8 kHz at 2 GHz carrier frequency.

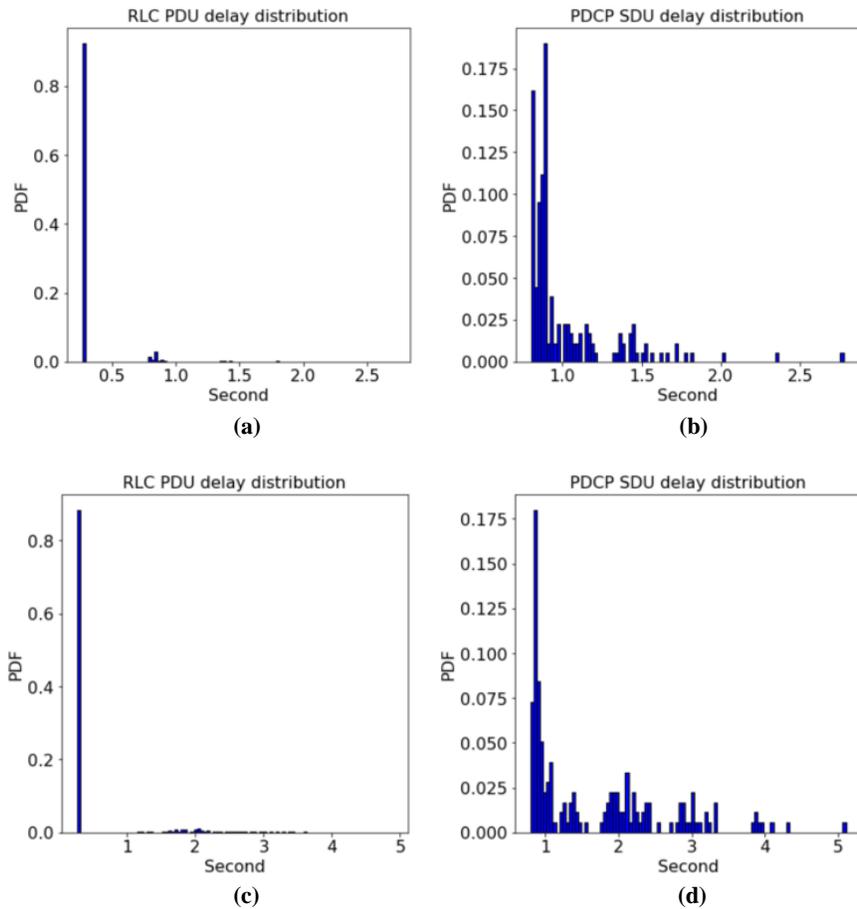

**Figure 5: Example simulated delay distributions with periodic packets (1 kilobyte per second), 256 ms one-way propagation delay, RLC acknowledged mode, and good link quality: subfigures (a) and (b) respectively show the delay distributions of RLC PDU and PDCP SDU when HARQ is used; subfigures (c) and (d) respectively show the delay distributions of RLC PDU and PDCP SDU when HARQ is not used.**

For non-GNSS based frequency adjustment techniques, the gNB may estimate the return link frequency shift of each UE and transmits a corresponding frequency adjustment command to the UE. To establish the uplink orthogonality as early as possible, it is desirable that the gNB estimates the uplink frequency shift from the random-access preamble transmitted by the UE and includes the frequency adjustment command in the random-access response message. The existing PRACH formats in 5G NR are however designed to facilitate uplink timing estimation and may need to be further enhanced to facilitate uplink frequency estimation.

### C. Hybrid Automatic Repeat Request

To combat against transmission errors, 5G NR uses a combination of forward error correction and automatic repeat request (ARQ), which is known as hybrid ARQ (HARQ). Forward error correction coding adds parity bits to the information bits, enabling error correction at the receiver. The receiver sends a positive acknowledgement if no error is detected in the received packet and a negative acknowledgement otherwise. In case of retransmission, the receiver can use soft combing to combine the retransmission with earlier transmission(s) for decoding. With incremental redundancy, each retransmission including additional parity bits progressively reduces the code rate, and thus HARQ may be viewed as an implicit link adaptation scheme.

5G NR supports 16 HARQ processes with stop-and-wait protocols per component carrier in both uplink and downlink. In a stop-and-wait protocol, the transmitter stops and waits for acknowledgement after each (re)transmission. Using 16 HARQ processes with stop-and-wait protocols would lead to significant throughput reduction especially in GEO communications systems.

One straightforward approach is to increase the number of HARQ processes to cope with the increased round-trip delays in satellite communications systems. This, however, comes at the cost of UE implementation complexity due to the increased UE HARQ soft buffer size. Another approach is to introduce a mechanism in 5G NR to support the possibility of turning off retransmissions in HARQ processes. Instead, the retransmissions are handled by the layers above MAC if error-free data units are required at the receiver. For example, the radio link control (RLC) layer supports an acknowledged mode that may be used for retransmissions of erroneous data. Retransmissions at the layers above MAC might lead to increased latency due to the slower feedback. Additionally, the communications system may need to operate with more conservative coding rate in the physical layer to avoid excessive retransmissions in the layers above MAC.

Figure 5 shows example simulated delay distributions with periodic packets (1 kilobyte per second), a one-way propagation delay of 256 ms, RLC acknowledged mode, and good link quality. The delay of RLC protocol data units (PDU) includes the delays in physical layer and MAC layer. The delay of packet data convergence protocol (PDCP) service data units (SDU) includes the delays in physical layer, MAC layer, RLC layer, and PDCP layer. Figures 5 (a) and 5 (b) respectively show the delay distributions of RLC PDU and PDCP SDU when HARQ is used and the transport block size is 1000 bits. Due to the good link quality, most of the RLC PDUs are successfully received without HARQ retransmission and thus have ~256 ms delays, while the remaining small fraction of RLC PDUs are successfully received with one HARQ retransmission and thus have ~800 ms delays. Accordingly, most of the PDCP SDUs have delays in the range of 0.8 s - 2 s. In contrast, when HARQ is not used, Figures 5 (c) and 5 (d) respectively show the delay distributions of RLC PDU and PDCP SDU. While most of the RLC PDUs have ~256 ms delays due to the good link quality, the remaining small fraction of RLC PDUs have delays greater than ~1.25 s. Accordingly, compared to the case with HARQ, more PDCP SDUs have delays greater than ~2 s when HARQ is not used. This is because HARQ feedback is faster than feedback in higher layer protocols.

### D. Idle Mode UE Tracking and Paging

When the UE is in connected mode, the network fully controls the serving cell of the UE. If the UE moves, the network initiates handover for the UE. However, when the UE is in idle mode, it is up to the UE on which cell it is camping as long as it follows specified rules for the cell selection and reselection.

Paging is the mechanism that the network uses to initiate a connection with a UE in idle mode. While camping on a cell, the UE wakes up at certain periods to monitor paging information from the network. The location of a UE in idle mode is known to the network at tracking area level. A tracking area is a cluster of gNBs or cells that an operator can define. Tracking areas in Release-15 NR are geographically nonoverlapping. The network can provide a UE with a list of tracking areas where the UE registration is valid. When the network pages the UE, the paging request can be transmitted to all the gNBs in the tracking area list, and then each gNB broadcasts the paging message in its cell. Therefore, the network needs to track the UE in idle mode at tracking area level, and this is achieved by periodic tracking area updates performed by the UE. In particular, when a UE in idle mode selects or reselects a cell, it reads the broadcasted system information to learn which tracking area this cell belongs to. If the cell does not belong to at least one of the tracking areas to which the UE is registered to, it performs a tracking area update to notify the network of the tracking area of the cell it is currently camping on.

For a GEO satellite communications system where the cell's coverage area is usually fixed on the ground, the existing UE tracking and paging procedures in 5G NR can be largely reused. However, for a non-GEO satellite communications system particularly with moving beams, the cell's coverage area moves on the ground. Under the existing UE tracking and paging procedures in 5G NR designed for terrestrial networks, the tracking area sweeps over the ground as well. As a result, a stationary UE would have to keep performing location registration in idle mode. How to efficiently perform UE tracking and paging in this case requires further thinking.

## VI. CONCLUSIONS

It is an interesting time to witness the comeback of satellite communications. The on-going evolution of 5G standards provides a unique opportunity to revisit satellite communications. Though 5G NR has been designed mainly targeting terrestrial mobile communications, the inherent flexibility of 5G NR allows it to be evolved to support non-terrestrial communications. As this article has highlighted, when adapting 5G NR to support satellite communications, there are challenges including long propagation delays, large Doppler effects, and moving cells. Addressing such challenges requires a rethinking of many of the working assumptions and models used to date for designing 5G NR. Throughout this article, we have attempted to highlight ideas on how to overcome the key technical challenges faced by 5G NR evolution for satellite communications.

## ACKNOWLEDGEMENT

The authors thank Ali Khayrallah for his valuable comments and suggestions.